\documentclass[twocolumn,pra,showpacs]{revtex4}

\begin{document}

\title{Atomic Clocks and the Search for Variation of the Fine Structure Constant}

\author{E.J. Angstmann}
\affiliation{School of Physics, University of New South Wales,
Sydney 2052, Australia}
\author{V. A. Dzuba}
\affiliation{School of Physics, University of New South Wales,
Sydney 2052, Australia}
\author{V.V. Flambaum}
\affiliation{School of Physics, University of New South Wales,
Sydney 2052, Australia}
\date{\today}

\begin{abstract}
Evidence that the fine structure constant, $\alpha = e^{2}/\hbar c$, was different in an early epoch has recently been found in quasar (QSO) absorption spectra. An accurate laboratory measurement of $\alpha$ variation could help to distinguish between space and time variation of $\alpha$ and possibly confirm the QSO results. We have performed calculations of the relativistic energy shifts that a variation of $\alpha$ would cause in a wide range of atomic spectra. A laboratory measurement of how these transition frequencies vary over time is vital in determining whether $\alpha$ is varying today.
\end{abstract}

\pacs{06.20.Jr, 31.30.Jv}
\maketitle

\section{Introduction}

Currently there is a lot of debate about whether the fine structure constant, $\alpha$, is varying. Several theories including Kaluza-Klein theories and string theories allow or even require $\alpha$ variation. These theories are higher-dimensional theories, our four dimensional constants depend upon the value of some scalar fields and on the structure and sizes of the extra dimensions. Any evolution of the size of the higher dimensions with time could lead to a spacetime variation of our four dimensional constants. An excellent review by Uzan \cite{Uzan} describes the theories that predict spacetime variation of constants as well as current limits on variation of these constants. There is evidence that other constants are also varying, for example it has been found that the deuteron binding energy has changed since the big bang \cite{Dmitriev}. Confirmation of $\alpha$ variation would be exciting since it would indicate new physics beyond the standard model.

Various methods have been applied to obtain a limit on $\alpha$ variation. Different methods span different time periods and so a comparison of the different constraints will lead to a picture of when and if $\alpha$ has varied. One of these methods involves the Okolo uranium mine, which is situated in Gabon in west Africa. It contained a natural fission reactor that was active 1.8 million years ago. The present isotopic abundances allow the reaction rates, while the fission reactor was active, to be extracted. This in turn allows a bound on changes in $\alpha$ to be extracted. The most recent limit obtained in this way is $\Delta \alpha/\alpha \ge 4.5 \times 10^{-8}$  ($6\sigma$ confidence)\cite{Lamoreaux}, where $\Delta \alpha$ is the change in $\alpha$ since the fission reactor was active, this corresponds to a redshift, $z \sim 0.14$. One of the problems with a limit on $\alpha$ variation obtained in this way is that there are a number of assumptions made in deriving the value.

Another method is to analyse QSO spectra to see whether $\alpha$ was different in the past. The big advantage of using QSOs is that they are situated at relatively large redshifts, $z \sim 0.5 - 3.5$ and so studying them allows us to peer into the ancient history of the universe. This is a big advantage since if $\alpha$ was varying steadily through time we would expect to see the largest change in the oldest spectral lines. Alternatively, if $\alpha$ varied suddenly, then since this data spans a larger percentage of the universe's history, the time at which $\alpha$ varied is more likely to be contained within this data. Several groups are currently working on obtaining limits on $\alpha$ variation from QSO spectra. There is some discrepancy among the results obtained by the different groups. The current limits on $\alpha$ obtained in this manner are shown in the Table \ref{tab:QSO}. 
\begin{table*}
\caption{Limits on $\alpha$ variation from QSO spectra.}
\begin{ruledtabular}
\begin{tabular}{c c c c}
\label{tab:QSO}
Reference & $\Delta \alpha/\alpha$ & Redshift & Technique\footnote{MM stands for the many-multiplet technique and RMM stands for regressional many-multiplet technique described in \cite{Quast}.} \\
\hline
Murphy \textit{et al. }\cite{Murphy} & $(-0.574\pm0.102) \times 10^{-5}$ & $0.2<z<3.7$ & MM\\
Quast \textit{et al. }\cite{Quast} & $(-0.04 \pm 0.19 \pm 0.27_{sys}) \times 10^{-5}$ & $1.15$ & RMM \\
Srianand \textit{et al. }\cite{Srianand} & $(-0.06 \pm 0.06) \times 10^{-5}$ & $0.4<z<2.3$ & MM \\
\end{tabular}
\end{ruledtabular}
\end{table*}

Several atomic clock type experiments have placed limits on present day $\alpha$ variation. These experiments do not yet match the precision of the QSO limits. The most precise limit obtained to date is $|\dot{\alpha}/\alpha|<1.2\times 10^{-15}/$yr. This was obtained by Bize \textit{et al.} \cite{Bize} using a frequency comb with a Hg$^{+}$ optical clock and a Cs microwave clock. It has been suggested by Nguyen \textit{et al.} \cite{Nguyen} that a sensitivity of $|\dot{\alpha}/\alpha|<1.2\times 10^{-18}/$yr could be reached by using the accidentally degenerate levels in Dysprosium, as suggested in \cite{Dzuba}. More laboratory experiments are vital in order to place stringent limits on present day $\alpha$ variation. Laboratory experiments are very sensitive to oscillatory variation of $\alpha$, this may be missed using other methods that involve comparing the values of $\alpha$ over a longer time period.
 
\section{Theory}

The basic concept behind using atomic clocks to measure a variation in $\alpha$ is that different atomic transitions depend differently upon $\alpha$. Comparing the rates of different atomic clocks over long periods of time allows one to put bounds on the local change of $\alpha$ with time. In deciding which atomic transitions to use several factors need to be taken into account; these factors include the lifetime of the level (and hence the width of the level), and the size of the relativistic effects in the level. An ideal situation would be to have two atomic clocks, each with a very narrow transition width, with very different relativistic effects. If $\alpha$ was varying these levels would then drift either apart or towards each other and this drift should be measurable over a long enough period of time.

We have performed calculations to determine how fast atomic levels will move apart in the presence of $\alpha$ variation. To do this accurate relativistic calculations are needed. It is convenient to represent the energy of a level by
\begin{equation}
\omega = \omega_{0} +qx
\end{equation}
where $x = (\alpha/\alpha_{0})^{2}-1$, $\omega_{0}$ is the initial value of $\omega$ and $\alpha_{0}$ is the initial value of $\alpha$, $q$ is a coefficient that determines the frequency dependence on $\alpha$ variation. The value of the $q$ coefficients for many transitions of interest for an atomic clock style experiment are presented in Table \ref{tab:qcoeff}. Ideally one would use two levels with very different $q$ coefficients. If the $q$ coefficients have opposite sign the levels will drift in opposite directions. Alternatively if you choose a level with a small $q$ coefficient this could act as an anchor against which the movement of another level could be measured.   

To find the value of the $q$ coefficients we have repeated calculations of the energy levels for different values of $\alpha$. We start the calculations from the relativistic Hartree-Fock (also known as Dirac-Hartree-Fock) method. We then use the combination of the configuration interaction (CI) method with the many-body perturbation theory (MBPT) \cite{Dzuba96,Dzuba98}. Interactions between valence electrons are treated using the CI method while correlations between the valence electrons and the core electrons are included by means of MBPT. For more details of the calculations see the references cited in Table \ref{tab:qcoeff}.

\begin{table}
\caption{Experimental energies and calculated $q$ coefficients for transitions from the ground state to the state shown.}  
\begin{ruledtabular}
\begin{tabular}{c c c c c c c}
\label{tab:qcoeff}
Atom/Ion & Z & \multicolumn{2}{c}{State} & Wavelength, \AA & q (cm$^{-1}$) & Reference\\
 & & & & Experiment & &\\
\hline
Al II & 13 & $3s3p$ & $^{3}P_{0}$ & 2674.30 & 146 & \cite{Angstmann} \\
      &    & $3s3p$ & $^{3}P_{1}$ & 2669.95 & 211 & \cite{Angstmann} \\
      &    & $3s3p$ & $^{3}P_{2}$ & 2661.15 & 343 & \cite{Angstmann} \\
      &    & $3s3p$ & $^{1}P_{1}$ & 1670.79 & 278 & \cite{Angstmann} \\
Ca I  & 20 & $4s4p$ & $^{3}P_{0}$ & 6597.22 & 125 & \cite{Angstmann} \\
      &    & $4s4p$ & $^{3}P_{1}$ & 6574.60 & 180 & \cite{Angstmann} \\
      &    & $4s4p$ & $^{3}P_{2}$ & 6529.15 & 294 & \cite{Angstmann} \\
      &    & $4s4p$ & $^{1}P_{1}$ & 4227.92 & 250 & \cite{Angstmann} \\
Sr I  & 38 & $5s5p$ & $^{3}P_{0}$ & 6984.45 & 443 & \cite{Angstmann} \\
      &    & $5s5p$ & $^{3}P_{1}$ & 6894.48 & 642 & \cite{Angstmann} \\
      &    & $5s5p$ & $^{3}P_{2}$ & 6712.06 & 1084 & \cite{Angstmann} \\
      &    & $5s5p$ & $^{1}P_{1}$ & 4608.62 & 924 & \cite{Angstmann} \\
Sr II & 38 & $4d$ & $^{2}D_{3/2}$ & 6870.07 & 2828 & \cite{Dzuba} \\
      &    & $4d$ & $^{2}D_{5/2}$ & 6740.25 & 3172 & \cite{Dzuba} \\
In II & 49 & $5s5p$ & $^{3}P_{0}$ & 2365.46 & 3787 & \cite{Angstmann} \\
      &    & $5s5p$ & $^{3}P_{1}$ & 2306.86 & 4860 & \cite{Angstmann} \\
      &    & $5s5p$ & $^{3}P_{2}$ & 2182.12 & 7767 & \cite{Angstmann} \\
      &    & $5s5p$ & $^{1}P_{1}$ & 1586.45 & 6467 & \cite{Angstmann} \\
Ba II & 56 & $5d$ & $^{2}D_{3/2}$ & 20644.74 & 5844 & \cite{Dzuba00} \\
      &    & $5d$ & $^{2}D_{5/2}$ & 17621.70 & 5976 & \cite{Dzuba00} \\
Dy I  & 66 & $4f^{10}5d6s$ & $^{3}[10]_{10}$ & 5051.03 & 6008 & \cite{Dzuba03}\\
      &    & $4f^{9}5d^{2}6s$ & $^{9}K_{10}$ & 5051.03 & -23708& \cite{Dzuba03} \\
Yb I  & 70 & $6s6p$ & $^{3}P_{0}$ & 5784.21 & 2714 & \cite{Angstmann} \\
      &    & $6s6p$ & $^{3}P_{1}$ & 5558.02 & 3527 & \cite{Angstmann} \\
      &    & $6s6p$ & $^{3}P_{2}$ & 5073.47 & 5883 & \cite{Angstmann} \\
      &    & $6s6p$ & $^{1}P_{1}$ & 3989.11 & 4951 & \cite{Angstmann} \\
Yb II & 70 & $4f^{14}5d$ & $^{2}D_{3/2}$ & 4355.25 & 10118 & \cite{Dzuba03} \\
      &    & $4f^{14}5d$ & $^{2}D_{5/2}$ & 4109.70 & 10397 & \cite{Dzuba03} \\
      &    & $4f^{13}6S^{2}$ & $^{2}F_{7/2}$ & 4668.81 & -56737 & \cite{Dzuba03} \\
Yb III& 70 & $4f^{13}5d$ & $^{3}P_{0}$ & 2208.63 & -27800 & \cite{Dzuba03} \\
Hg I  & 80 & $6s6p$ & $^{3}P_{0}$ & 2656.39 & 15299 & \cite{Angstmann} \\
      &    & $6s6p$ & $^{3}P_{1}$ & 2537.28 & 17584 & \cite{Angstmann} \\
      &    & $6s6p$ & $^{3}P_{2}$ & 2270.51 & 24908 & \cite{Angstmann} \\
      &    & $6s6p$ & $^{1}P_{1}$ & 1849.50 & 22789 & \cite{Angstmann} \\
Hg II & 80 & $5d^{9}6s^{2}$ & $^{2}D_{5/2}$ & 2815.79 & -56671 & \cite{Dzuba} \\
      &    & $5d^{9}6s^{2}$ & $^{2}D_{3/2}$ & 1978.16 & -44003 & \cite{Dzuba} \\
Tl II & 81 & $6s6p$ & $^{3}P_{0}$ & 2022.20 & 16267 & \cite{Angstmann} \\
      &    & $6s6p$ & $^{3}P_{1}$ & 1872.90 & 18845 & \cite{Angstmann} \\
      &    & $6s6p$ & $^{3}P_{2}$ & 1620.09 & 33268 & \cite{Angstmann} \\
      &    & $6s6p$ & $^{1}P_{1}$ & 1322.75 & 29418 & \cite{Angstmann} \\
Ra II & 88 & $6d$ & $^{2}D_{3/2}$ & 8275.15 & 18785 & \cite{Dzuba00} \\
      &    & $6d$ & $^{2}D_{5/2}$ & 7276.37 & 17941 & \cite{Dzuba00} \\
\end{tabular}
\end{ruledtabular}
\end{table}

\section{Conclusion}
Obtaining limits on the variation of fundamental constants is a very active and exciting topic of research since it can indicate new physics beyond the standard model. Experiments need to be conducted to place limits on the variation of constants. To measure the variation of $\alpha$ in a laboratory the relative movement of two atomic energy levels with different $q$ coefficients needs to be precisely measured over a period of time. Note that if one measures absolute frequency variation using the Cs hyperfine frequency standard then the dependence of the Cs frequency on $\alpha$, calculated in \cite{Dzuba96}, and on $(m_{q}/\Lambda_{QCD})$ where $m_{q}$ is the quark mass and $\Lambda_{QCD}$ is the strong interaction (QCD) scale, calculated in \cite{Flambaum} must be taken into account. 

\section*{Acknowledgment}
This work was supported by the Australian Research Council.

\end{document}